# Peptide strings clues to the genesis and treatment of rheumatoid arthritis: rebuilding self-protective immunity amid fungal ruins

**Razvan Tudor Radulescu**

Molecular Concepts Research (MCR), Muenster, Germany

E-mail: ratura@gmx.net

Mottos:   *"...nature may, after all, be entirely approachable. Her much-advertised inscrutability has once more found to be an illusion due to our ignorance. This is encouraging, for if the world in which we live were as complicated as some of our friends would have us believe, we might well despair that biology could ever become an exact science."*
(Thomas Hunt Morgan)

*Simplex sigillium veri.*

## ABSTRACT

A recent application of the peptide strings concept has yielded novel perceptions on cell growth regulation, for instance that of oncoprotein metastasis. Here, this interdisciplinary approach at the boundary between physics and biology has been applied to gain a more profound insight into rheumatoid arthritis. As a result of the present investigation, this disease could be viewed as due to a metabolic dysregulation/syndrome-associated breakdown in the immunoglobulin A-based surveillance of the potentially pathogenic fungus *Candida albicans* that subsequently engenders a widespread self-destruction through cross-reactive auto-epitopes, ultimately amounting to the systemic predominance of a pro-inflammatory peptide string. Its therapeutic counterpart equally proposed in this report might serve as a model for future strategies against autoimmunity.

About 2 years ago, Weyand and Goronzy summarized the molecular complexity of rheumatoid arthritis (RA) and emphasized that its seemingly disparate aspects are still in need of a basic explanation, thereby calling for a sort of (physical) string theory to reconcile the various pathophysiologic details of this autoimmune disease (1). This appeal prompted me to explore whether my interdisciplinary concepts on defensology (2) and peptide strings (3-5) could yield some useful insights towards a more simplified and fundamental view on RA.

Since one of the hallmarks of RA is the so-called rheumatoid factor (RF), i.e. an autoantibody directed against the constant region of IgG, i.e. immunoglobulin G (6), I first addressed the issue as to whether the tetrapeptide fragment Thr-Lys-Val-Asp (TKVD) that I had discovered (2) to be shared by the constant region of IgG1 and the retinoblastoma tumor suppressor protein (RB) could be part of an immunoregulatory peptide string which may have become deregulated in RA.

Peptide strings are an emergent property of biologically important proteins/peptides whereby these molecules display the following features (4,5):

a)  related to each other in that they share a distinct amino acid motif (associated with a common biological function);
b)  besides the above functional agonism as defined in a) and depending on a given cellular context, equally capable of functional antagonism and/or of binding to one another (as a result of molecular complementariness);
c)  capable of self-binding and/or containing functionally antagonistic domains/fragments whereby one of the two counterparts carries the common motif defined in a);
d)  able to translocate across relatively large subcellular distances (e.g. from the cell nucleus/cytosol to the extracellular space and/or vice versa);
e)  if considered together as one functional unit, found, by abstraction, to trace a transcellular route across cells.

Briefly, a peptide string can emerge from functional links between an extracellular protein, a cytoplasmic protein and a nuclear protein, any of which is able to translocate to one another´s compartment and to bind each other as well as to self-associate through common amino acid motifs, as initially illustrated for oncogenic and anti-oncogenic processes (4).

The minimum requirement for a peptide string, however, is probably even less: a self-binding protein with functionally antagonistic domains as well as dual or triple (extracellular/cytoplasmic/nuclear) localization in which either of 2 possible antagonistic peptide strings can be "imprinted" depending on the nature of the environmental/hormonal cue to which it both binds and resembles (qualitative aspect) and, moreover, by the amount of such interactions over spacetime (quantitative aspect). For instance, a growth-inhibitory RB peptide string corresponds to relatively few insulin-RB complex formations across one or more cells whereas a growth-promoting RB peptide string emerges when such insulin-bound RB fraction outnumbers the amount of free RB (or yet E2F-bound) molecules (5). Interestingly, this latter RB string is well-suited to explain the recently observed oncogenic implication of RB in ras-induced transformation processes (7,8). Likewise, insulin-driven oncoprotein metastasis (9,10) can be perceived in terms of a persistently active growth-promoting (RB) peptide string which, in its mechanistic essence, may equally account for certain forms of chemical carcinogenesis for which a "protein deletion" mechanism had been proposed in the late 1940s (11). Therefore, it is likely that the degree of neutralization of an antineoplastic RB peptide information over spacetime would decide on the ultimate growth-inhibitory or growth-promoting nature of the corresponding RB peptide string.

In the light of this underlying concept, one possible TKVD-based peptide string could be formed by extracellular and cytoplasmic IgG1 along with cytoplasmic and nuclear RB whereby the biological information encoded in the TKVD tetrapeptide would be passed on- similar to the baton within a relay race- primarily from internalized, cytoplasmic IgG1 to cytoplasmic RB through direct physical interaction and/or indirectly by the intermediary of another cytoplasmic (and nuclear) protein.

Such intermediary could be the- cytoplasmic and nuclear (12)- clathrin heavy chain (CHC) whereby its type 1 isoform (CHC1) intriguingly harbors the TKVD tetrapeptide, as I have now found (notably, the type 2 CHC contains the highly related SKVD tetrapeptide, respectively, and could thus also serve as a functional link between IgG1 and RB). On the one hand, such possibility of an association between IgG1 and CHC is supported by a study showing clathrin-dependent IgG1 endocytosis (13) and, on the other hand, CHC may also contact RB given the close cytoplasmic proximity of the RB-binding protein p600 with clathrin (14).

Interestingly, I have identified the TKVD peptide also in the Rac1 protein of the (extra- and intracellular) potentially pathogenic fungus *Candida albicans* (*C. albicans*), yet not in its human homologue (15).

This finding suggested that an uncontrolled *C. albicans* infection may interfere in susceptible individuals with an TKVD-based immunoregulatory string through molecular mimicry of such (physiological) peptide string and, by engendering recognition of the TKVD self-epitopes on RB, CHC and IgG1 as cross-reactive epitopes, contribute to the autoimmune pathogenesis of RA. Remarkably, this hypothesis of mine is supported by the well-known association between (chronic) candidiasis and autoimmunity (16,17) as well as by the fact that *C. albicans* components can elicit an RA-like disease in experimental animals, albeit the chosen fungal immunogen was a ß-glucan, not a protein fragment in this setting (18).

Yet, why would this deleterious cascade be initiated and proceed in some individuals and not be suppressed by anti-fungal defense mechanisms at an early stage? The potential answer likely resides in the peptide string normally opposing or rather counter-balancing the TKVD-based peptide string as part of a homeostatic process. Such anti-peptide string needs to be surmised as being deficient or dysfunctional in order for the TKVD-based peptide string to predominate in a pathologic manner.

Before, however, further discussing this function, its structural details have to be defined. Based on the above peptide strings definition, the peptide string functionally opposite to the transcellularly conveyed TKVD amino acid information could be defined in a straightforward manner if a (self)binding site for TKVD was known.

In the absence of such knowledge, I therefore had to choose the equally possible alternative of identifying such anti-peptide string by searching for proteins/peptides known to antagonize/attenuate pathologic aspects of RA. My search yielded calcitonin as a possible candidate given that this hormone has been shown to have beneficial effects in the treatment of RA (19,20) which may be due to its previously demonstrated ability to inhibit macrophage function (21). Under the premise that calcitonin may contain peptide information overlapping with an existing anti-arthritogenic peptide string that involves RB and CHC under physiological circumstances, I performed a sequence comparison of all these molecules in the form of tripeptide sequences. This analysis yielded that RB and calcitonin share only the 2 tripeptides Gly-Asn-Leu (GNL) and Asn-Leu-Ser (NLS) whereas CHC and calcitonin share the 7 tripeptides GNL, LST, TQD, QTA, TAI, IGV and VGA.

Moreover, to achieve a perfect functional mirroring to the TKVD-based peptide string, an immunoglobulin had to be searched for that possibly shares with RB and/or CHC as well as calcitonin one of the above tripeptides. Following this insight, my additional exploration has yielded that, remarkably, the variable region of a distinct human immunoglobulin A1 heavy chain, briefly IgA1 $V_H$ (22), bears the NLS tripeptide whereas a comparable domain within the human IgG1 heavy chain (23) does not contain any of the above tripeptides. Notably, CHC harbors (in either of its two isoforms) the highly related sequence N*F*S.

Along the same lines, I have detected the NLS sequence also in CD16 (i.e. a receptor specifically recognizing the constant region of its ligand IgG, briefly an Fc receptor), thus supporting the possibility of a direct physical interaction between TKVD (e.g. in IgG) and NLS (e.g. in CD16) peptide sequences, besides their above anticipated involvement in mediating opposite biological effects.

Since such found NLS sequences conform with the universal glycosylation motif NXS whereby X signifies any amino acid, the following intriguing scenario is conceivable. Normally, the NLS sequence in IgA1 $V_H$ may fullfill 2 roles:

a) to antagonize Candida infection by neutralizing the TKVD peptide in the Rac1 protein of this fungus;

b) to mask the TKVD epi/idiotope in IgG1 such that it is not recognized as an auto-antigenic site. As far as this second point is concerned and expressed differently, I assume that physiologically there are some IgA1 anti-idiotypic antibodies that, as part of an anti-arthritogenic idiotype network, suppress the "surfacing" of potentially autoimmune sites in (the constant region of) IgG1. Such host-beneficial masking of this IgG1 site would probably be performed not only by IgA1, but also by CD16.

In this context, it should be specified that the TKVD tetrapeptide is located in close proximity to the IgG hinge region whose conformational change presumably accounts for the structurally altered serum IgG of RA patients as compared to that of healthy individuals (24,25), thus supporting the possibility that IgG1 TKVD may be unmasked as an auto-epitope for RF. Consistent with this prediction, Johnson and coworkers had presciently supposed that "new determinants are exposed at or near the hinge region of the rheumatoid IgG molecule" (26). Notably, the investigations carried out by Johnson *et al.* (25,26) have remained particularly important since they were performed with RA patient-derived IgG whereas subsequent (crystallization or binding) studies were primarily conducted with artificially produced IgG (Fc fragment).

Alternatively or additionally, the TKVD sequence could be self-targeted as part of an IgG fragment proteolytically generated during the disease process given that IgG degradation is a relevant aspect of RA pathogenesis (27,28).

By contrast, if the above NLS site in IgA1 was abnormally glycosylated, this may impair both anti-microbial/anti-fungal defense (29) and the prevention of autoimmunity. The latter would be likewise affected in case the CD16 NLS sequence was also abnormally glycosylated.

My present scenario is supported by the following facts. Firstly, it has been shown that (the onset of) RA coincides with aberrant glycosylation processes (30) that may affect not only IgG (31), but also other plasma proteins (32). Such phenomenon is complemented by the observation according to which this disease is also characterized by an insulin resistance along with a glucose utilization defect (33,34), ultimately promoting the emergence of AGE, i.e. advanced glycation endproducts (35), among which some may directly correlate with the activity of RA (36).

Moreover and intriguingly, my present assumption on a causative IgA defect in RA is in full accordance with the well-known, yet poorly understood tendency of IgA-deficient individuals to develop immune-complex disease (37) and, more generally viewed, with the increasingly appreciated link between immunodeficiency and autoimmunity (38,39).

Since, furthermore, *C. albicans* Rac1 also contains the NLS peptide besides the TKVD peptide, it could be presumed that these peptide sequences are involved in the self-binding and folding of this protein. If, by contrast, (excessive) glycosylation occurred at its NLS site, the *C. albicans* Rac1 TKVD may be unmasked. Given that (diabetes mellitus-associated) hyperglycemic states are known to be associated with an increased frequency of fungal infections, it could be anticipated that such unmasked TKVD site contributes to the augmented virulence of these fungi under such conditions.

Consequently, I would assume that at the root of RA there is a (metabolically induced, specifically AGE-associated) dysfunction in *C. albicans*-specific as well as anti-IgG1 anti-idiotypic IgA1 in the blood circulation, on the skin and/or on mucosal surfaces.

As a result, *Candida* fungi could then invade host cells and spread throughout the body in an unchecked manner and, moreover, the TKVD epitope in IgG1, CHC and RB would be unmasked and thus become an autoimmune epitope due to the interference of the identical TKVD sequence in the Rac1 protein of *C. albicans* with (host-protective) NLS peptide-like domains in IgA1, CD16, CHC and RB. Ultimately, an anti-*Candida* immune response would be mounted that, due to the above mentioned antigenic similarity, is also directed against IgG1, CHC and RB. The putative self-neutralization of these 3 targets by RF(s) would conceivably engender the following molecular consequences:

a) through its autoimmune destruction, IgG may no longer be able to auto-regulate its own expression (40,41), thus exacerbating the concurrent IgA defect-based immunodepression towards specific foreign antigens and thus paving the way to a state of uncontrolled inflammation;

b) CHC self-elimination would entrain a lack of termination of inflammation-associated signalling since normally pro-inflammatory receptors such as TLR-4 are silenced by clathrin-associated endocytic recycling, thus further enhancing the hyper-inflammatory mechanism described in a);

c) RB self-targeting should be similarly detrimental as RB is likely to be a key conductor in the physiological repression of the inflammatory response (2) and thus its abolition would parallel its known inactivation by inflammatory mediators such as NF-kB, especially during hyper-inflammatory conditions.

In keeping with this third point, it has been shown that RF cross-reacts with IgG and nuclear antigens the latter of which are not only histones (42) and, furthermore, that a subset of RA patients possesses antinuclear antibodies (ANA) directed against 105 and 95 kD antigens (43). These 2 autoantigens are conceivably RB isoforms, the former with its predominant molecular weight (44) and the latter potentially being a dysfunctional deletion variant as seen in human osteosarcoma cells (45). Moreover, a likely RA-preventive role of RB is also suggested by the fact that p21, i.e. one of its primary activators and inducers, inhibits interleukin-6 and matrix metalloproteinase-1 (46) since such p21 effect may well be mediated by RB similar to other p21-associated phenomena such as the cell cycle arrest induced by gamma-irradiation (47,48).

If this scenario held true, then immunization of experimental animals with the TKVD peptide should cause an RA-like disease in them.

Figs. 1a and 1b (cf. page 12) sketch the basic principles of dissemination of a given peptide information in the sense of a peptide string.

Tables 1-3 (cf. page 13) summarize this peptide strings view on normal TKVD- as well as NLS peptide-based immunoregulation vs. early and advanced forms of RA ("hum." is the abbreviation for "human", the term "amino acid" has been abridged as "aa", square brackets indicate diminished function, crossed lines signify the destruction of the respectively specified protein, absence of an involved protein in a certain table box corresponds to its functional silencing and red color highlights the predominant peptide string).

Furthermore, Fig. 2 (cf. page 14) represents an application of the tenets of Fig. 1 to the contents of Tables 1-3 and, as such, displays how e.g. the propagation of the TKVD autoimmune peptide information could be imagined under primarily stereospecific or steric considerations.

Last but not least, Fig. 3 (cf. page 15) represents a further abstraction and simplification of the same scenario as it focuses on capturing it in the form of two opposite facets of an immunoregulatory peptide string for RB- depicted in a graphical form that has previously been introduced to illustrate growth-regulatory RB peptide strings (5)- whereby the NLS-based (anti-inflammatory, anti-arthritogenic) RB string predominates under physiological conditions (Fig. 3A) and, by contrast, the TKVD-based (pro-inflammatory/arthritogenic/autoimmune) RB string prevails during RA (Fig. 3B). Thus, in the former case, it is the NLS sequence which, in analogy to Mendel´s genetic traits, could be considered as functionally "dominant" over the functionally "recessive" TKVD whereas, in the latter case, these features would be reversed.

Consequently, a possible future treatment of early-stage RA could include *C. albicans*-specific anti-fungal agents.

Yet, in order to catch up with a potential autoimmune cascade of events already initiated by *C. albicans*, the present peptide strings interpretation of RA suggests that it may be necessary to administer a peptide comprising the above Asn-Leu-Ser or Asn-Phe-Ser sequence, briefly an NL/FS amino acid signature, in pharmacological doses in order to mask the potentially arthritogenic TKVD tetrapeptide such that it is no longer visible for an overreacting immune system and hence to curtail the inflammatory response against this antigen by restoring the natural balance or homeostasis between TKVD and NL/FS peptide strings which, similar to the predominance of growth-inhibitory peptide strings in the normal state (5), would be asymmetric such that anti-inflammatory NL/FS peptide strings would prevail to a certain degree.

Moreover, the NL/FS peptide-based "epitope masking therapy" I am proposing here would have to be administered systemically and such that it reaches the inside of all cells in the body.

The reason therefore is that peptide strings, whether agonistic or antagonistic, describe a generalized transcellular condition with resemblance (5) to the resonance thought to govern chemical bonds the latter of which is an arbitrary and idealized, yet valuable quantum mechanical approximation (49). Hence, a pathogenic peptide strings condition could only be reversed by an opposite, likewise generalized (also resonance-like) *state* that is not limited to a single antigenic molecule or distinct target cell, but instead recruits a common peptide motif in various proteins within and across many different cells (4,5).

Should future studies choose to experimentally explore, besides the previously reported growth-regulatory peptide strings (3-5) and their corrolaries (9,10), also this peptide string scenario for RA and its therapeutic counterpart, there is a fair chance that not only a fundamentally new signal transmission mode- partly resembling the (free radical) chain reactions described by Semenov (50)- will be further understood, but that additionally Szent-Györgyi´s bold and long-standing expectation of quantum-mechanical explanations for otherwise elusive biological processes (51) will finally be specifically addressed along the way.

## References

1. Weyand, C.M., and Goronzy, J.J. 2006. Pathomechanisms in rheumatoid arthritis - time for a string theory? *J. Clin. Invest.* **116:**869-871.
2. Radulescu, R.T. 2006. Defensology: a universal theory of host defense against microbial infection, inflammation and cancer involving retinoblastoma protein (RB). *Int. Med.* **1:**12-17.
3. Radulescu, R.T. 2005. From particle biology to protein and peptide strings: a new perception of life at the nanoscale. *Logical Biol.* **5:**98-100.
4. Radulescu, R.T. 2006. Peptide strings in detail: first paradigm for the theory of everything (TOE). *Pioneer* **1:**62-68.
5. Radulescu, R.T. 2007. Across and beyond the cell are peptide strings. *arXiv*:0711.0202v1 [q-bio.SC].
6. Egeland, T., and Munthe, E. 1983. The role of the laboratory in rheumatology. Rheumatoid factors. *Clin. Rheum. Dis.* **9:**135-160.
7. Williams, J.P., et al. 2006. The retinoblastoma protein is required for ras-induced oncogenic transformation. *Mol. Cell. Biol.* **26:**1170-1182.
8. Degregori, J. 2006. Surprising dependency for retinoblastoma protein in ras-mediated tumorigenesis. *Mol. Cell. Biol.* **26:**1165-1169.
9. Radulescu, R.T. 2007. Oncoprotein metastasis disjoined. *arXiv*:0712.2981v1 [q-bio.SC].
10. Radulescu, R.T. 2008. Going beyond the genetic view of cancer. *Proc. Natl. Acad. Sci. USA* **105:**E12.
11. Miller, E.C., and Miller, J.A. 1947. The presence and significance of bound aminoazo dyes in the livers of rats fed p-dimethylaminoazobenzene. *Cancer Res.* **7:**468-480.
12. Enari, M., Ohmori, K., Kitabayashi, I., and Taya, Y. 2006. Requirement of clathrin heavy chain for p53-mediated transcription. *Genes Dev.* **20:**1087-1099.
13. Pearse, B.M.F. 1982. Coated vesicles from human placenta carry ferritin, transferrin, and immunoglobulin G. *Proc. Natl. Acad. Sci. USA* **79:**451-455.
14. Nakatani, Y., et al. 2005. p600, a unique protein required for membrane morphogenesis and cell survival. *Proc. Natl. Acad. Sci. USA* **102:**15093-15098.
15. Bassilana, M., and Arkowitz, R.A. 2006. Rac1 and Cdc42 have different roles in *Candida albicans* development. *Eukaryotic Cell* **5:**321-329.
16. Mathur, S., Melchers, J.T. 3rd, Ades, E.W., Williamson, H.O., and Fudenberg, H.H. 1980. Anti-ovarian and anti-lymphocyte antibodies in patients with chronic vaginal candidiasis. *J. Reprod. Immunol.* **2:**247-262.
17. Peterson, P., Pitkänen, J., Sillanpää, N., and Krohn, K. 2004. Autoimmune polyendocrinopathy candidiasis ectodermal dystrophy (APECED): a model disease to study molecular aspects of endocrine autoimmunity. *Clin. Exp. Immunol.* **135:**348-357.
18. Hida, S., Miura, N.N., Adachi, Y., and Ohno, N. 2007. Cell wall ß-glucan derived from Candida albicans acts as trigger for autoimmune arthritis in SKG mice. *Biol. Pharm. Bull.* **30:**1589-1592.
19. Sileghem, A., Geusens, P., and Dequeker, J. 1992. Intranasal calcitonin for the prevention of bone erosion and bone loss in rheumatoid arthritis. *Ann. Rheum. Dis.* **51:**761-764.
20. Kröger, H, Arnala, I., and Alhava, E.M. 1992. Effect of calcitonin on bone histomorphometry and bone metabolism in rheumatoid arthritis. *Calcif. Tissue Int.* **50:**11-13.
21. Nong, Y.H., Titus, R.G., Ribeiro, J.M., and Remold, H.G. 1989. Peptides encoded by the calcitonin gene inhibit macrophage function. *J. Immunol.* **143:**45-49.

22. Putnam, F.W., Liu, Y.-S.V., and Low, T.L.K. 1979. Primary structure of a human IgA1 immunoglobulin. *J. Biol. Chem.* **254:**2865-2874.
23. NCBI protein database accession code AAF03881.
24. Watkins, J., Unger, A., and Mahon, N. 1970. Differences between papain hydrolysis patterns of the serum gammaG-globulin of healthy individuals and of rheumatoid patients. *Clin. Sci.* **38:**15P-16P.
25. Johnson, P.M., Watkins, J., Scopes, P.M., and Tracey, B.M. 1974. Differences in serum IgG structure in health and rheumatoid disease. Circular dichroism studies. *Ann. Rheum. Dis.* **33:**366-370.
26. Johnson, P.M., Papamichail, M., Gutierrez, C., and Holborow, E.J. 1975. Interaction of the hinge region of human immunoglobulin G with a murine lymphocyte membrane receptor. Relevance to the problem of antiglobulin induction in rheumatoid arthritis. *Immunology* **28:**797-805.
27. Watkins, J., Roberts, A., and Johnson, P.M. 1975. Catabolism of human IgG in mice sensitized to various IgG fragments. Similarities to the catabolism of rheumatoid IgG in mice. *Immunology* **28:**755-759.
28. Ryan, M.H. et al. 2008. Proteolysis of purified IgGs by human and bacterial enzymes in vitro and the detection of specific proteolytic fragments of endogenous IgG in rheumatoid synovial fluid. *Mol. Immunol.* **45:**1837-1846.
29. Mathur, S., et al. 1977. Humoral immunity in vaginal candidiasis. *Infect. Immun.* **15:**287-294.
30. Parekh, R.B., Dwek, R.A., and Rademacher, T.W. 1988. Rheumatoid arthritis as a glycosylation disorder. *Br. J. Rheumatol.* 27 Suppl. **2:**162-169.
31. Parekh, R.B., et al. 1985. Association of rheumatoid arthritis and primary osteoarthritis with changes in the glycosylation pattern of total serum IgG. *Nature* **316:**452-457.
32. Raghav, S.K. et al. 2006. Altered expression and glycosylation of plasma proteins in rheumatoid arthritis. *Glycoconj. J.* **23:**167-173.
33. Liefmann, R. 1949. Endocrine imbalance in rheumatoid arthritis and rheumatoid spondylits; hyperglycemia unresponsiveness, insulin resistance, increased gluconeogenesis and mesenchymal tissue degeneration; preliminary report. *Acta Med. Scand.* **136:**226-232.
34. Paolisso, G., et al. 1991. Evidence for peripheral impaired glucose handling in patients with connective tissue diseases. *Metabolism* **40:**902-907.
35. Soldatos, G., Cooper, M.E., and Jandeleit-Dahm, K.A. 2005. Advanced-glycation end products in insulin-resistant states. *Curr. Hypertens. Rep.* **7:**96-102.
36. Chen, J.R., et al. 1998. Pentosidine in synovial fluid in osteoarthritis and rheumatoid arthritis: relationship with disease activity in rheumatoid arthritis. *J. Rheumatol.* **25:**2440-2444.
37. Roitt, I., Brostoff, J., and Male, D. 1996. Immunology. *Mosby*, 4th edition, p. 21.2.
38. Notarangelo, L.D., Gambinieri, E., and Badolato, R. 2006. Immunodeficiencies with autoimmune consequences. *Adv. Immunol.* **89:**321-370.
39. Coutinho, A., and Carneiro-Sampaio, M. 2008. Primary immunodeficiencies unravel critical aspects of the pathophysiology of autoimmunity and of the genetics of autoimmune disease. *J. Clin. Immunol.* 28 Suppl. **1:**S4-S10.
40. Radulescu, R.T. 1995. Antibody constant region: potential to bind metal and nucleic acid. *Med. Hypotheses* **44:**137-145.
41. Radulescu, R.T. 1998. Immune modulation by zinc: clues from immunoglobulin structure and function. *Immunol. Today* **19:**288.
42. Hobbs, R.N., Lea, D.J., Phua, K.K., and Johnson, P.M. 1983. Binding of isolated rheumatoid factors to histone proteins and basic polycations. *Ann. Rheum. Dis.* **42:**435-438.

43. Labrador, M., et al. 1998. Antibodies against a novel nucleolar and cytoplasmic antigen (p105-p42) present in the sera of patients with a subset of rheumatoid arthritis (RA) with signs of scleroderma. *Clin. Exp. Immunol.* **114:**301-310.
44. Maxwell, S.A. 1994. Retinoblastoma protein in non-small cell lung carcinoma cells arrested for growth by retinoic acid. *Anticancer Res.* **14:**1535-1540.
45. Shew, J.-Y., et al. 1990. C-terminal truncation of the retinoblastoma gene product leads to functional inactivation. *Proc. Natl. Acad. Sci. USA* **87:**6-10.
46. Perlman, H., et al. 2003. IL-6 and matrix metalloproteinase-1 are regulated by the cyclin-dependent kinase inhibitor p21 in synovial fibroblasts. *J. Immunol.* **170:**838-845.
47. Harrington, E.A., Bruce, J.L., Harlow, E., and Dyson, N. 1998. pRB plays an essential role in cell cycle arrest induced by DNA damage. *Proc. Natl. Acad. Sci. USA* **95:**11945-11950.
48. Brugarolas, J., et al. 1999. Inhibition of cyclin-dependent kinase 2 by p21 is necessary for retinoblastoma protein-mediated G1 arrest after gamma-irradiation. *Proc. Natl. Acad. Sci. USA* **96:**1002-1007.
49. Pauling, L. 1960. The nature of the chemical bond. *Cornell University Press*, 3rd edition, pp. 13-14.
50. Semenov, N.N. 1956. Some problems relating to chain reactions and to the theory of combustion. *Nobel Lecture*, pp. 487-514.
51. Szent-Györgyi, A. 1941. Towards a new biochemistry? *Science* **93:**609-611.

**Fig. 1**

**a)**

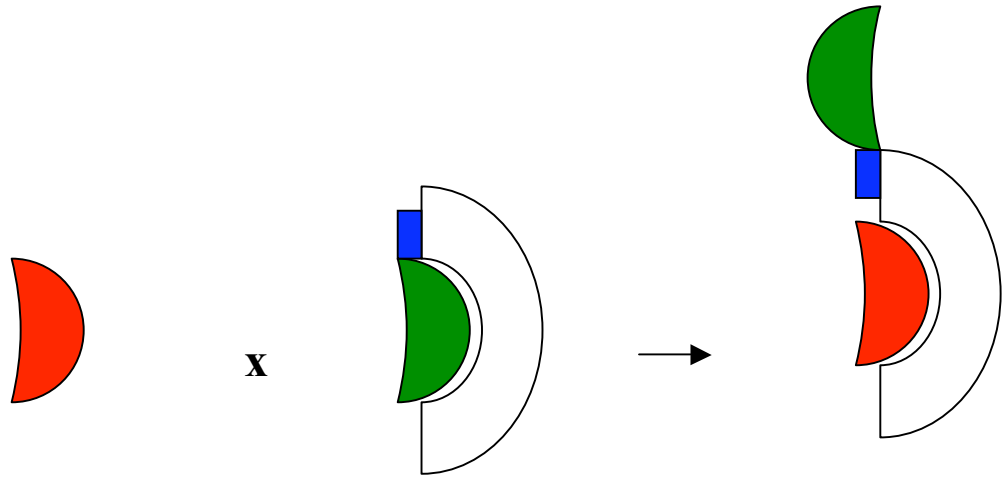

**b)**

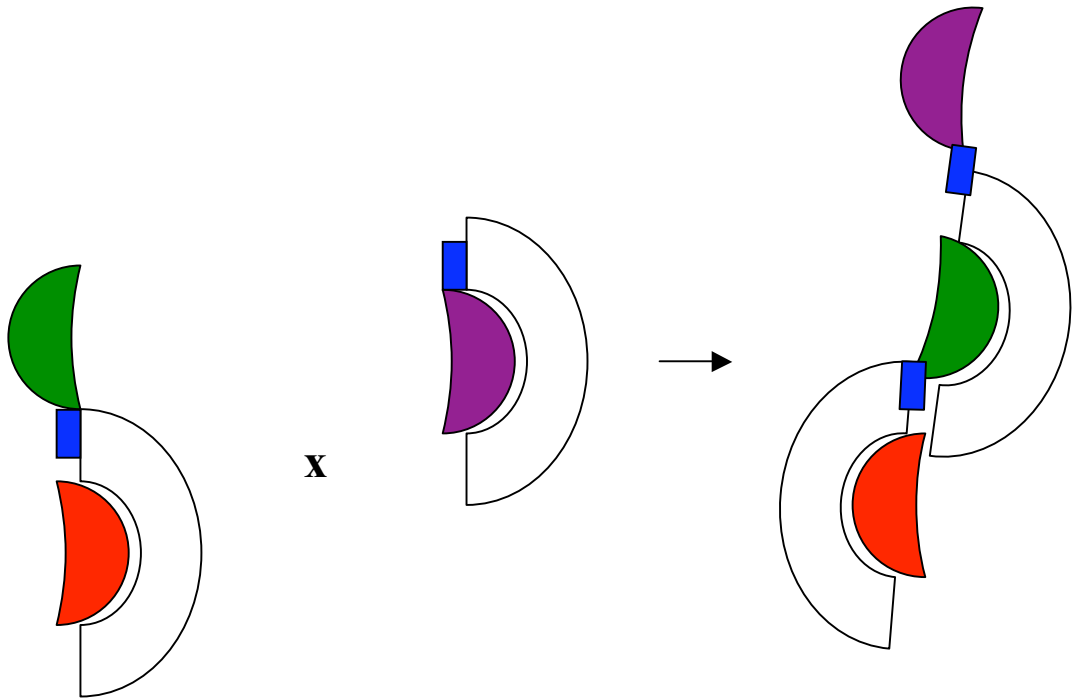

Fig. 1a   Peptide information transfer from an outer cue (red) to a bipartite, self-complementary/self-binding protein with a subdomain (green) resembling the cue whereby such subdomain would be sterically or allosterically exposed upon binding of the cue to such protein, thus initiating a peptide string.

Fig. 1b   Peptide information transfer from the protein specified in a) to another bipartite, self-complementary/self-binding protein with a subdomain (purple) resembling both the cue and the cue-like subdomain in the protein initially bound by the cue whereby in the course of this protein-protein interaction such subdomain in the second protein would be sterically or allosterically exposed, thus further propagating or disseminating the peptide string initiated by the cue.

**Tab. 1**

| NORMAL | Arthritogenic peptide string | Anti-arthritogenic peptide string |
|---|---|---|
| Extracellular environment | Thr-Lys-Val-Asp  hum.IgG1 C$_H$ (aa 209-212) | Asn-Leu-Ser  hum.IgA1 V$_H$ (aa 28-30)<br>Asn-Leu-Ser  hum.CD16 (aa 71-73)<br>Asn-Leu-Ser  hum.calcitonin (aa 3-5) |
| Cytoplasm | Thr-Lys-Val-Asp  hum.IgG1 C$_H$ (aa 209-212)<br>Thr-Lys-Val-Asp  hum.CHC1 (aa 1608-1611)<br>Thr-Lys-Val-Asp  hum.RB (aa 142-145) | Asn-Leu-Ser  hum.IgA1 V$_H$ (aa 28-30)<br>Asn-*Phe*-Ser  hum.CHC1 (aa 720-722)<br>Asn-Leu-Ser  hum.RB (aa 316-318) |
| Nucleus | Thr-Lys-Val-Asp  hum.CHC1 (aa 1608-1611)<br>Thr-Lys-Val-Asp  hum.RB (aa 142-145) | Asn-*Phe*-Ser  hum.CHC1 (aa 720-722)<br>Asn-Leu-Ser  hum.RB (aa 316-318) |

**Tab. 2**

| EARLY RHEUMATOID ARTHRITIS | Arthritogenic peptide string | Anti-arthritogenic peptide string |
|---|---|---|
| Extracellular environment | Thr-Lys-Val-Asp  C.albicans Rac1 (aa 116-119)<br>Thr-Lys-Val-Asp  hum.IgG1 C$_H$ (aa 209-212) | [Asn-Leu-Ser  hum.IgA1 V$_H$ (aa 28-30)]<br>[Asn-Leu-Ser  hum.CD16 (aa 71-73)]<br>[Asn-Leu-Ser  hum.calcitonin (aa 3-5)] |
| Cytoplasm | Thr-Lys-Val-Asp  C.albicans Rac1 (aa 116-119)<br>Thr-Lys-Val-Asp  hum.IgG1 C$_H$ (aa 209-212)<br>Thr-Lys-Val-Asp  hum.CHC1 (aa 1608-1611)<br>Thr-Lys-Val-Asp  hum.RB (aa 142-145) | [Asn-Leu-Ser  hum.IgA1 V$_H$ (aa 28-30)]<br>[Asn-*Phe*-Ser  hum.CHC1 (aa 720-722)]<br>[Asn-Leu-Ser  hum.RB (aa 316-318)] |
| Nucleus | Thr-Lys-Val-Asp  hum.CHC1 (aa 1608-1611)<br>Thr-Lys-Val-Asp  hum.RB (aa 142-145) | [Asn-*Phe*-Ser  hum.CHC1 (aa 720-722)]<br>[Asn-Leu-Ser  hum.RB (aa 316-318)] |

**Tab. 3**

| ADVANCED RHEUMATOID ARTHRITIS | Arthritogenic peptide string | Anti-arthritogenic peptide string |
|---|---|---|
| Extracellular environment | [Thr-Lys-Val-Asp  C.albicans Rac1 (aa 116-119)]<br>~~Thr-Lys-Val-Asp  hum.IgG1 C$_H$ (aa 209-212)~~ | |
| Cytoplasm | [Thr-Lys-Val-Asp  C.albicans Rac1 (aa 116-119)]<br>~~Thr-Lys-Val-Asp  hum.IgG1 C$_H$ (aa 209-212)~~<br>~~Thr-Lys-Val-Asp  hum.CHC1 (aa 1608-1611)~~<br>~~Thr-Lys-Val-Asp  hum.RB (aa 142-145)~~ | ~~Asn-*Phe*-Ser  hum.CHC1 (aa 720-722)~~<br>~~Asn-Leu-Ser  hum.RB (aa 316-318)~~ |
| Nucleus | ~~Thr-Lys-Val-Asp  hum.CHC1 (aa 1608-1611)~~<br>~~Thr-Lys-Val-Asp  hum.RB (aa 142-145)~~ | ~~Asn-*Phe*-Ser  hum.CHC1 (aa 720-722)~~<br>~~Asn-Leu-Ser  hum.RB (aa 316-318)~~ |

**Fig. 2**

**a)**

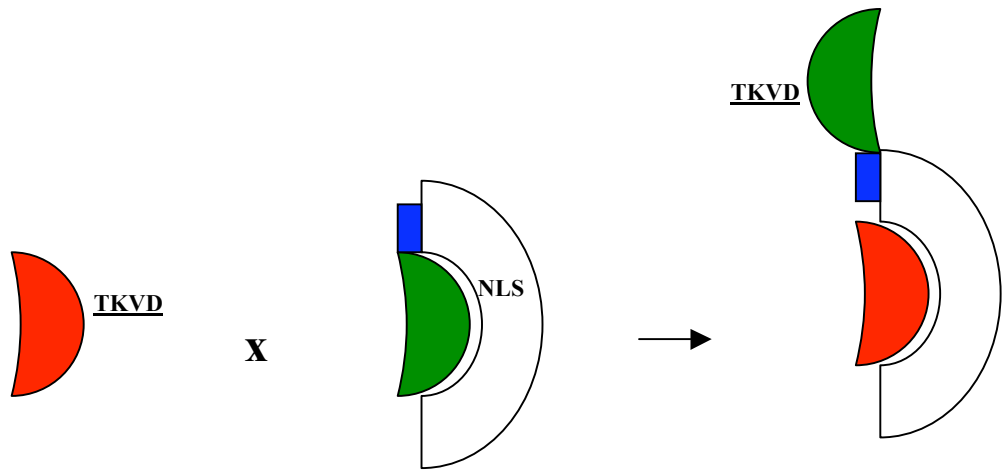

**b)**

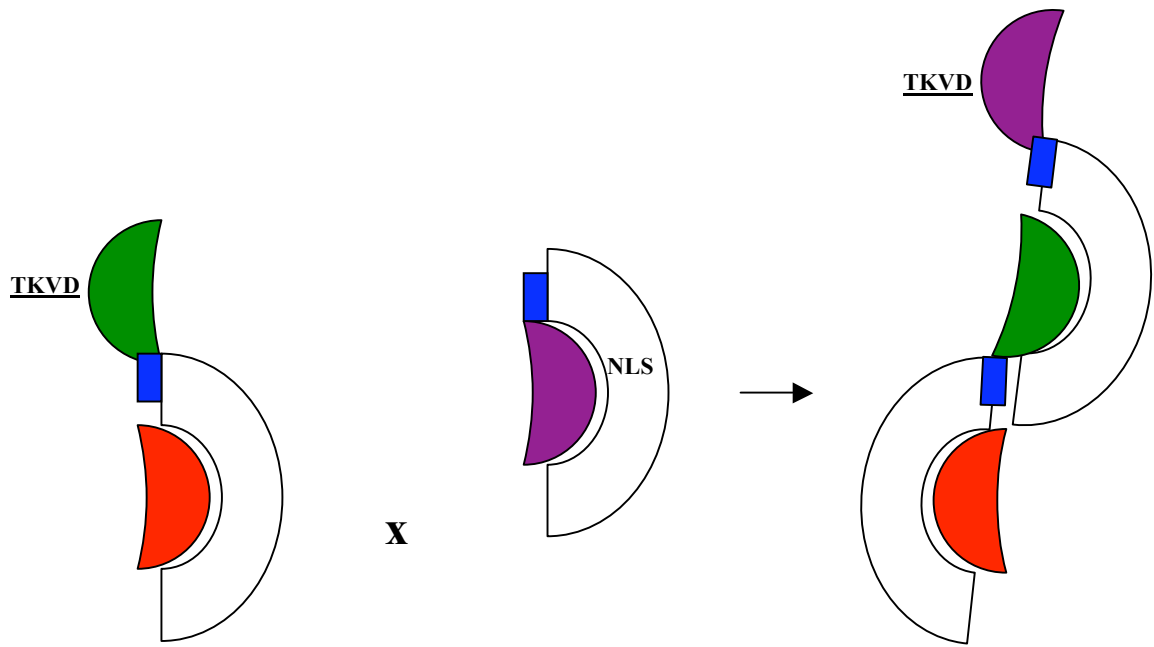

Fig. 2a   Peptide information transfer from an outer cue carrying the TKVD amino acid sequence (red) to a bipartite, self-complementary/self-binding protein with a subdomain (green) harboring the same tetrapeptide whereby such subdomain would be sterically or allosterically exposed upon binding of the cue to another subdomain in such protein containing the NLS tripeptide, thus initiating a TKVD peptide string.

Fig. 2b   Peptide information transfer from the "cue-activated" protein specified in a) to another bipartite, self-complementary/self-binding protein comprising both the TKVD sequence (purple) and the NLS tripeptide whereby the TKVD tetrapeptide of the second protein would be sterically or allosterically exposed upon this protein-protein interaction, thus further propagating or disseminating the TKVD peptide string initiated by the cue.

**Fig. 3**

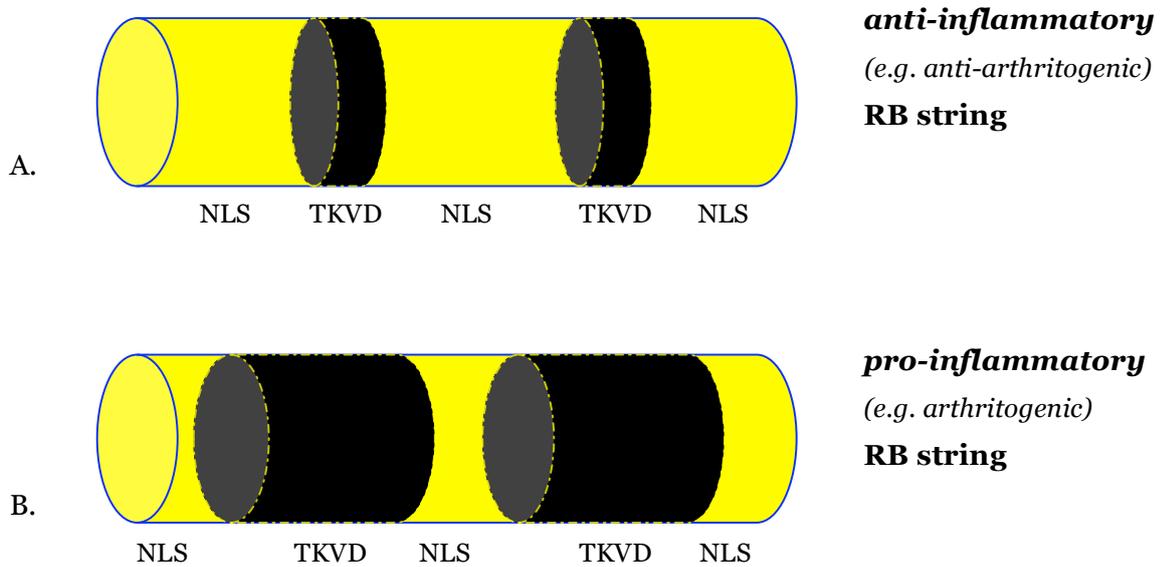

A.  NLS    TKVD    NLS    TKVD    NLS

*anti-inflammatory*
*(e.g. anti-arthritogenic)*
**RB string**

B.  NLS    TKVD    NLS    TKVD    NLS

*pro-inflammatory*
*(e.g. arthritogenic)*
**RB string**

Fig. 3A   anti-inflammatory peptide string for retinoblastoma protein (RB) corresponds to a majority of RB molecules with an exposed Asn-Leu-Ser, briefly NLS, fragment and a minority of RB molecules with an exposed Thr-Lys-Val-Asp, briefly TKVD, fragment in a certain spacetime;

Fig. 3B   pro-inflammatory peptide string for RB corresponds to a majority of RB molecules with an exposed Thr-Lys-Val-Asp, briefly TKVD, fragment and a minority of RB molecules with an exposed Asn-Leu-Ser, briefly NLS, fragment in a certain spacetime.